\documentclass[twocolumn]{article}
\usepackage{amsmath}

\def\annexes{
	\setcounter{section}{0}\def\thesection{\Alph{section}}	
	\numberwithin{equation}{section}}

\def\betas{\beta^{*}}
\def\Ht{\hat H}
\def\betat{\hat \beta}

\def\qa{q_\alpha}
\def\ha{h_\alpha}
\def\ra{r_\alpha}
\def\xa{x_\alpha}
\def\qb{q_\beta}
\def\xb{x_\beta}
\def\a{\alpha}
\def\b{\beta}
\def\r#1{(\ref{#1})}
\def\bin#1#2{\left(#1\atop#2\right)}

\title{Probability distribution of the free energy of a directed 
polymer in a random medium}

\author{\'Eric Brunet\quad and\quad Bernard Derrida\\[2mm]
  \small \emph{Laboratoire de Physique Statistique,}\\[-1.2mm] 
  \small \emph{\'Ecole Normale Sup\'erieure,}\\[-1.2mm]
  \small \emph{24, rue Lhomond,}\\[-1.2mm] 
  \small \emph{75231 Paris C\'edex 05, France.}\\[2mm]
  \small \texttt{Eric.Brunet@physique.ens.fr}\\[-1.2mm]
  \small \texttt{Bernard.Derrida@physique.ens.fr}\\[3mm]
  \small Physical Review E \textbf{61} (6), (2000).}

\begin{document}
\twocolumn[%
\maketitle
\begin{abstract}
We calculate exactly the first cumulants
of the free energy of a
directed polymer in a random medium for the geometry of a cylinder. By
using the fact that the $n$-th moment $\langle Z^n\rangle$ of the partition
function is given by the ground state energy of a quantum problem of $n$
interacting particles on a ring of length $L$, we write an integral
equation allowing to expand these moments in powers of the strength of
the disorder $\gamma$ or in powers of $n$. For $n$ small and $n \sim (L
\gamma)^{-1/2}$, the moments~$\langle Z^n\rangle$ take a
scaling form which allows to describe all the fluctuations of order $1/L$
 of the free energy per unit length of the
directed polymer. The distribution of these fluctuations is the same as the
one found recently in the asymmetric exclusion process, indicating that it
is characteristic of all the systems described by the Kardar-Parisi-Zhang
equation in~$1+1$ dimensions.

\end{abstract}\bigskip
P.A.C.S. numbers: 05.30, 05.70, 64.60\,Cn.\bigskip]

\section{Introduction}

Directed polymers in a random medium is one of the simplest systems for
which the effect of strong disorder can be
studied\cite{KardarZhang.87,Zhang.87,Halpin-HealyZhang.95}. At the mean
field level, it possesses a low temperature phase, with a broken symmetry
of replica\cite{DerridaSpohn.88,Derrida.91} similar to mean field spin
glasses\cite{MezardParisiVirasoro.87}. The problem is however much better
understood than spin glasses; in particular one can
write\cite{DerridaSpohn.88,Derrida.91} closed expressions of the mean field
free energy and one can predict the existence\cite{ImbrieSpencer.88} of
phase transitions in all dimensions~$d+1 > 2+1$. It is also an interesting
system from the point of view of non-equilibrium phenomena: through the
Kardar-Parisi-Zhang (KPZ) equation\cite{KardarParisiZhang.86,Krug.97}, it
is related to ballistic growth models and, in $1+1$ dimensions, to the
asymmetric simple exclusion process
(ASEP)\cite{Halpin-HealyZhang.95,Krug.97}.

In the theory of disordered systems, the replica approach plays a very
special role. On the one hand, it is one of the most powerful theoretical
tools and often the only possible approach to study some strongly
disordered systems. On the other hand it is difficult to tell in advance
whether the predictions of the replica approach are correct or not. When
it does not work, one can always try to break the symmetry of
replica\cite{MezardParisiVirasoro.87}: this usually makes the calculations
much more complicated without being certain that the results
become correct. 
In the replica approach, the calculation usually starts with an integer
number $n$ of replica. Then, as the limit of physical interest is the
limit $n \to 0$, one has to extend to non-integer~$n$ results obtained for
integer $n$. This is in fact the big difficulty of the replica approach, so
it is useful to look at
simple examples for which the $n$ dependence can be studied in detail.

This is one of the motivations of the present work, where we show how to
calculate integer and non-integer moments $\langle Z^n \rangle$ of the
partition function $Z$ of a directed polymer in $1+1$ dimensions. The
geometry we consider is a cylinder infinite in the $t$ direction and
periodic, of size $L$, in the $x$ direction (\textit{i.e.} $x +L \equiv
x$). The partition function $Z(x,t)$ of a directed polymer joining the
points $(0,0)$ and $(x,t)$ on this cylinder is given by the path integral
\begin{align}
Z(x,t) = \int_{(0,0)}^{(x,t)} \kern-1.5em {\cal D} y(s) \exp \Bigg( -&
\int_0^t \kern-0.7em ds\, \bigg[ {1 \over 2 } \left( d y(s) \over ds
\right)^2 + \nonumber\\
&\eta(y(s),s) \bigg] \Bigg) ,
\label{path}
\end{align}
where the random medium is characterised by a Gaussian white
noise~$\eta(x,t)$
\begin{equation}
\langle \eta(x,t) \eta(x',t') \rangle = \gamma \ \delta( x-x') \
\delta(t-t'). \label{noise}
\end{equation}
One of the main goals of the present work is to calculate
the cumulants $\lim_{t\to\infty} \langle\ln^k Z(t)\rangle_c/t$ of
the free energy per unit length of the directed polymer. These
cumulants are the coefficients of the small $n$ expansion
of~$E(n,L,\gamma)$ defined as 
\begin{equation}
E(n,L,\gamma)= -
\lim_{t \to \infty} {1 \over t} \ln \left[ \langle Z^n(x,t) \rangle \over
\langle Z(x,t) \rangle^n \right].
\label{defE}
\end{equation}
This~$E(n,L,\gamma)$ was calculated exactly by Kardar\cite{Kardar.87} for
integer~$n$ and~$L=\infty$. His closed
expression~$E(n,\infty,\gamma)=-n(n^2-1)\gamma^2/24$ cannot however be
continued to all values of~$n$, in particular to negative~$n$, as it would
violate the fact that $\partial^2 E(n,L,\gamma)/\partial n^2$ is negative. 
Therefore one does not know the range of validity of this expression.

The second motivation of the present work is to test the universality class
of the KPZ equation. The problem~\r{path} of a directed polymer in a
random medium is described by the KPZ equation as several other
problems such as growing interfaces or exclusion
processes\cite{Halpin-HealyZhang.95}. For certain models of this class,
the asymmetric exclusion processes, the distribution of the total
current $Y_t$ integrated over time~$t$ has been calculated
exactly\cite{DerridaLebowitz.98,DerridaAppert.99,Kim.95,LeeKim.99,LebowitzSpohn.99}
in the long time limit. For large $t$, the generating function of this
integrated current $Y_t$ on a ring of $L$ sites takes the
form\cite{DerridaLebowitz.98,DerridaAppert.99}
\begin{equation}
\ln\langle e^{\alpha Y_t}\rangle \sim {\Lambda_{\text{max}}(\alpha) t},
\label{relYl}
\end{equation}
and it was
shown\cite{DerridaLebowitz.98,DerridaAppert.99,Kim.95,LeeKim.99},
when $L$ is large and when the parameter $\alpha$ in \r{relYl} is of
order $L^{-3/2}$, that~$\Lambda_{\text{max}}(\alpha)$ takes the following
scaling form
\begin{equation}
\label{scaling}
\Lambda_{\text{max}}(\alpha)-\alpha K_1 =K_2 G(\alpha K_3)
\end{equation}
where $K_1$, $K_2$ and $K_3$ are three constants which depend
on the system size $L$, the density of particles and the asymmetry.

The interesting aspect of \r{scaling} is that the function $G(\beta)$ is
universal\cite{DerridaAppert.99,LeeKim.99,Appert.00} in the sense that it
does not depend on any of the microscopic parameters which define the
model. It is given (in a parametric form) by
\begin{align}
\label{largedev}
\beta 	&= -\sum_{p=1}^{+\infty}{\epsilon^p \over p^{3/2}},\\
\label{largedev2}
G(\beta)&= -\sum_{p=1}^{+\infty}{\epsilon^p \over p^{5/2}}.
\end{align}

In the correspondence\cite{Halpin-HealyZhang.95} between the directed
polymer problem and the asymmetric exclusion process through the KPZ
equation, the role played by $\ln(Z(t))$ is the ratio $Y_t / L$. Comparing
$\langle \exp(\alpha Y_t ) \rangle $ and~$ \langle Z^n(t) \rangle$ in
equations~(\ref{defE}, \ref{relYl}), we see that~$n$ corresponds to~$\alpha
L$ and~$ E(n,L,\gamma)$ to~$\Lambda_{\text{max}}(\alpha)$. If the function
$G(\beta)$ is characteristic of systems described by the KPZ equation, we
expect in the scaling regime (large $L$ and $n \sim L^{-1/2}$), a relation
similar to~(\ref{scaling}) between $ E(n,L,\gamma) $ (defined by~\r{defE})
and $n$.  This is indeed one of the main results of the present work: when
$L$ is large and $n \sim L^{-1/2}$, we find
\begin{equation}
E(n,L,\gamma)={n\gamma^2\over24}-{\sqrt\gamma\over2\sqrt{2\pi}L^{3/2}}
G(-n\sqrt{2\pi L\gamma}). 
\label{mainresult} 
\end{equation} 
It is clear that in order to establish this relation we have to calculate
non-integer moments of the partition function.

The paper is organised as follows. In section~\ref{Schrodinger}, we recall
how the replica approach of~\r{path} can be formulated as a quantum problem
with~$n$ particles on a ring and how this problem can be solved by the
Bethe ansatz when the noise is $\delta $ correlated as in~(\ref{noise}). In
section~\ref{keyeq}, we write an integral equation~\r{eqB} which, together
with some symmetry conditions~(\ref{B(1)}, \ref{paritybis}), allows to
solve the Bethe equations of section~\ref{Schrodinger}. The main advantage
of~\r{eqB} is that the strength~$c$ of the disorder (where~$c=\gamma L/2$)
and the number of replica appear as continuous parameters. We show how
expansions in powers of~$c$ or in powers of the number~$n$ of replica can
be obtained from this integral equation. In the expansion of the
energy~$E(n,L,\gamma)$ in powers of~$c$, all the coefficients are
polynomials in~$n$. This allows us to define~$E(n,L,\gamma)$ for a 
non-integer~$n$ at least perturbatively in~$c$. At the end of
section~\ref{keyeq}, we show how to generate a small~$n$ expansion which
solves the integral equation~(\ref{eqB}). We also give explicit expressions
up to order $n^3$ and we notice that in this small~$n$ expansion of the
energy, we have to deal with coefficients that are functions of~$c$ with a
zero radius of convergence. The content of sections~\ref{Schrodinger}
and~\ref{keyeq} is essentially a recall of a method developed in our
previous work\cite{BrunetDerrida.00}. In section~\ref{scalingregime}, we
show that the recursion of section~\ref{keyeq} which generates all the
terms of the small $n$ expansion simplifies greatly in the scaling regime
($c$ large and~$n \sim c^{-1/2}$) allowing to calculate all the terms of
the expansion and to establish~(\ref{mainresult}).

\section{A quantum system of $n$ particles with $\delta$ interactions}
\label{Schrodinger}

Let us start with a case slightly more general than~\r{noise} where the noise
$\eta(x,t)$ in (\ref{path}) is a Gaussian noise $\delta$-correlated in time
but with some given correlation $v$ in space
\begin{equation}
\langle \eta(x,t) \eta(x',t') \rangle = \gamma\ v(x-x')\ \delta(t-t').
\label{corr}
\end{equation}
If we consider the correlation function $\langle
Z(x_1,t)\,Z(x_2,t)\,\dots\,Z(x_n,t)
\rangle$ of the partition function $Z(x,t)$ at points~$x_1$, $x_2$,\ldots
, $x_n$, one can
check\cite{Halpin-HealyZhang.95} from~(\ref{path}, \ref{corr}) that it
satisfies 
\begin{align}
&{d \over dt}
 \langle Z(x_1,t) Z(x_2,t) \dots Z(x_n,t) \rangle = \nonumber\\
&\qquad - \ {\cal \tilde{H} } \ 
 \langle Z(x_1,t) Z(x_2,t) \dots Z(x_n,t) \rangle .
\end{align}
where the Hamiltonian ${\cal \tilde{H}}$
is given by
\begin{equation}
{\cal \tilde{H}} = 
- {1\over2}\sum_\alpha {\partial^2\over\partial\xa^2}
-\gamma \sum_{\alpha<\beta} v(x_\alpha - x_\beta) - \gamma{n\over 2} v(0),
\label{tildeH}
\end{equation}
and where, because of the cylinder geometry in the directed polymer
problem, we have $x_\alpha\equiv x_\alpha+L$ for $1\le\alpha\le n$.

This implies that in the long time limit, 
\begin{equation}
 \langle Z(x_1,t) Z(x_2,t) \dots Z(x_n,t) \rangle \sim e^{-t {\tilde
 E}(n,L,\gamma)},
\end{equation}
where ${\tilde E}(n,L,\gamma)$ is the ground state energy of~(\ref{tildeH}).

If one takes the limit $v(x-x') \to \delta(x-x')$, the energy ${\tilde
E}(n,L,\gamma)$ becomes infinite because of the constant part $n v(0) /2$
in~\r{tildeH}. This divergence disappears, however, if we consider the
ratio $\langle Z(x_1,t) Z(x_2,t) ... Z(x_n,t) \rangle / \prod_\alpha
\langle Z(x_\alpha,t) \rangle$, and one can see that in the long time
limit,
\begin{equation}
 {\langle Z(x_1,t) Z(x_2,t) ... Z(x_n,t) \rangle \over 
\langle Z(x_1,t)\rangle \langle Z(x_2,t) \rangle ... \langle Z(x_n,t) \rangle} 
 \sim e^{-t E(n,L,\gamma)},
 \label{cumulants}
\end{equation}
where $E(n,L,\gamma)$ is the ground state energy of the Hamiltonian
\begin{equation}
{\cal H}=-{1\over2}\sum_\alpha {\partial^2\over\partial\xa^2}
		-\gamma \sum_{\alpha<\beta}
\delta(\xa-\xb),
\label{hamiltonian}
\end{equation}
where the positions $\xa$ of the $n$ particles are on a ring of length $L$.

Lieb and Liniger have shown that the Bethe ansatz allows to calculate the
ground state energy $ E(n,L,\gamma)$ of this one dimensional quantum
Hamiltonian exactly\cite{LiebLiniger.63,Lieb.63,YangYang.69,Gaudin.71,Jimboetall.80,Thacker.81,Gaudin.83}.
The Bethe ansatz consists in looking for a ground state wave function
$\Psi(x_1,\dots,x_n)$ of ~\r{hamiltonian} of the form
\begin{equation}
\Psi(x_1,\dots,x_n)=\sum_{P} a_P \ 
	e^{2(q_1 x_{P(1)}+\dots+q_n x_{P(n)})/L}
\label{defpsi}
\end{equation}
in the region $0\le x_1\le\dots\le x_n\le L$.
The sum in~\r{defpsi} runs over all the permutations $P$ of $\lbrace
1,\dots,n\rbrace$ and the value of $\Psi$ in other regions can be deduced
from (\ref{defpsi}) by symmetries. One can
show\cite{Jimboetall.80,Thacker.81,Gaudin.83,BrunetDerrida.00} that~\r{defpsi} is the ground
state wave function of~\r{hamiltonian} at energy 
\begin{equation}
 E(n,L,\gamma)=-{2\over L^2}\sum_{1\le\a\le n}\qa^2,
\label{defener}
\end{equation}
if the $\qa$ are the solutions of the~$n$ coupled equations
\begin{equation}
e^{2\qa}=\prod_{\b\ne\a}{\qa-\qb+{c}\over\qa-\qb-{c }},
\label{AnsatzSol}
\end{equation}
obtained by continuity from the solution $\lbrace\qa\rbrace = \lbrace 0
\rbrace$ at $c=0$ where 
\begin{equation}
c={\gamma L\over 2}.
\label{cdef}
\end{equation}
Moreover, the $q_\alpha$ are all different and the ground state is
symmetric~($\lbrace\qa\rbrace=\lbrace-\qa\rbrace$).
(See for instance \cite{Jimboetall.80}. Note that $i k_j$ and $c$
in~\cite{Jimboetall.80} are here ${2\over L}q_j$ and $-\gamma$; so our $c$
defined by \r{cdef} and the $c$ in~\cite{Jimboetall.80} are different.)

If we introduce the polynomial~$P(X)$ 
\begin{equation}
P(X) = \prod_{\qa} (X-\qa),
\label{defpoly}
\end{equation}
the system of equations~\r{AnsatzSol} becomes
\begin{equation}
e^{\qa} P(\qa-c) + e^{-\qa} P(\qa+c) =0,
\label{mainqa}
\end{equation}
for any $1\le\a\le n$, and we have from the symmetry of the ground state
\begin{equation}
P(-X)=(-1)^n P(X).
\label{parity}
\end{equation}
The knowledge of the polynomial $P(X)$ determines the energy~\r{defener}
as
\begin{equation}
P(X)=X^n -{1\over2}\left(\sum_{1\le\a\le n} \qa^2\right)X^{n-2} +\dots
\label{relPE}
\end{equation}
(using~\r{defpoly} and the fact that $\sum\qa=0$.)

For small $c$, it is possible to solve directly~\r{mainqa} and to determine
the $\qa$ (see appendix \ref{appDirect}). This leads to the following
expression of the ground state energy~(\ref{defener})
\begin{align}
\label{enc}
E(n,L,\gamma) = - {2 \over L^2} n(n-1)\left(
{c\over2}+{c^2\over12}+{nc^3\over180}+ O(c^4) \right).
\end{align}
We see that the first coefficients of the small~$c$ expansion are
polynomial in~$n$. In fact, following the approach of
appendix~\ref{appDirect}, one can see that
each coefficient of the small~$c$ expansion of~$E(n,L,\gamma)$
is polynomial in~$n$, allowing to define, at least perturbatively in~$c$,
the ground state energy~$E(n,L,\gamma)$ for non-integer~$n$.
The approach of appendix~\ref{appDirect} becomes however quickly complicated.
This is why in the next section we develop a different
approach\cite{BrunetDerrida.00} based on the integral equation~\r{eqB}.

\section{Solution of the Bethe ansatz using an integral equation}
\label{keyeq}

In this section we recall the approach developed in our previous
work\cite{BrunetDerrida.00}, which consists in writing an integral
equation where $c$ and $n$ appear as continuous parameters and which allows
to expand the energy in powers of $c$ as well as in powers of $n$.

Let us introduce the following function of~$\{ \qa \}$:
\begin{equation}
B(u)={1\over n}e^{c(u^2-1)/4}\sum_{\qa} \rho(\qa) e^{\qa(u-1)},
\label{defB}
\end{equation}
where the parameters $\rho(\qa)$ are defined by
\begin{equation}
\rho(\qa)=\prod_{\qb\ne\qa}{\qa-\qb+c\over \qa-\qb}.
\label{defrho}
\end{equation}
If the $\{ \qa \}$ are given by the solution of \r{AnsatzSol} which
corresponds to the ground state, one can show (see appendix
\ref{derivation}) that the function $B(u)$ satisfies the integral equation
\begin{align}
&B(1+u)-B(1-u) =\nonumber\\
&\qquad nc\int_0^u \kern-0.8em dv\,e^{-c(v^2-uv)/2}B(1-v)B(1+u-v).
\label{eqB}
\end{align}
and the following two conditions
\begin{align}
B(1)&=1,
\label{B(1)} \\
B(u)&=B(-u).
\label{paritybis}
\end{align}
Moreover, the energy (\ref{defener}) can be extracted from
the knowledge of $B(u)$ through 
\begin{equation}
\label{relEB}
E(n,L,\gamma) = {2 \over L^2} \left[{n^3c^2\over6}+{n c^2\over12}+{n
c\over2}
 - n B''(1) \right].
\end{equation}

The derivation of (\ref{eqB}, \ref{B(1)}, \ref{paritybis}, \ref{relEB}) is
given in appendix \ref{derivation}. We are now going to see how one can
find perturbatively in~$c$ or in~$n$ the solution of (\ref{eqB},
\ref{B(1)}, \ref{paritybis}) and, consequently, the ground state
energy~\r{relEB}.

\subsection{Expansion in powers of $c$}
\label{smallc}

To obtain the small $c$ expansion of $B(u)$ for arbitrary~$n$, we write
\begin{equation}
B(u)=B_0(u)+c B_1(u) +c^2 B_2(u)+\dots
\end{equation}
Conditions (\ref{B(1)}) and (\ref{paritybis}) impose that $B_0(0)=1$ and all
$B_k(1)=0$ for $k>0$, and that the $B_k(u)$ are all even. Moreover, as
can be seen directly from~\r{AnsatzSol}, the~$\qa$ scale like~$\sqrt c$
when~$c$ is small. (Appendix~\ref{appDirect} shows how to obtain the small~$c$
expansion of the~$\qa$.) This implies from the definition~\r{defB}
of~$B(u)$ that all the~$B_k(u)$ are polynomials in~$u$.

At zero-th order in $c$, \r{eqB} becomes:
\begin{equation}
B_0(1+u)-B_0(1-u)=0.
\label{eqB0}
\end{equation}
The only polynomial solution of~\r{eqB0} consistent with~(\ref{B(1)},
\ref{paritybis}), \textit{i.e.} $B_0(u)=B_0(-u)$ and $B_0(1)=1$ is simply
\begin{equation}
B_0(u)=1
\end{equation}
for any $u$. We put this back into~\r{eqB} and we get at first order in
$c$
\begin{equation}
B_1(1+u)-B_1(1-u)= n u.
\end{equation}
Again, there is a unique polynomial solution which satisfies the facts that
$B_1(u)$ is even and that $B_1(1)=0$:
\begin{equation}
B_1(u)={n\over4}(u^2-1).
\end{equation}
It is easy to see from (\ref{eqB})
that at any order in $c$, we have to solve 
\begin{align}
\label{genB}
B_k(1+u)-B_k(1-u)= \phi_k(u),
\end{align}
where $\phi_k(u)$ is
a polynomial odd in $u$. There is a unique even
polynomial $B_k(u)$ solution of~\r{genB} satisfying $B_k(1)=0$: it is one
degree higher than $\phi_k(u)$ and can be determined by equating each power
of~$u$ on both sides of~\r{genB}. $\big[$
Alternatively, we found a way of writing
the solution for any~$\phi_k(u)$:
\begin{align}
B_k(u)= &\Bigg[{s_0\over2} \int_1^u\kern-0.8emdv\, \phi_k(v) + {s_1\over2}
\left(\phi_k'(u)-\phi_k'(1)\right) \nonumber\\
 &+{s_2\over2} \left(\phi_k'''(u)- \phi_k'''(1)\right)+ \dots \nonumber\\
 &+{s_p\over2} \left(\phi_k^{(2p-1)} (u) - \phi_k^{(2p-1)} (1) \right) + \dots
\Bigg]/2\end{align}
where the $s_k$ are the coefficients of the expansion of $x/\sinh x$ in
powers of $x$ (\textit{i.e.} as $x/\sinh x = 1 - x^2 /6 + 7 x^4 / 360
+\dots$, one has $s_0=1$, $s_1= -1/6$, $s_2= 7 /360$, \ldots).\ $\big]$

This procedure gives for the first terms
\begin{align}
\label{Bexpn}
&B(u)=1 + { cn(u^2-1) \over 4} + 
 {c^2 n ( 2n + 1)( u^2-1)^2\over 96}\nonumber\\
&\quad+{c^3 n (  u^2 -1 )^2 
 \left(\displaystyle 5n^2 (  u^2 -1 )+4n(  2u^2 - 1 )
 \atop \displaystyle  + 2( u^2 - 3 ) \right) \over
	 5760}\nonumber\\
&\quad+O(c^4).
\end{align}
The energy can then be deduced from~\r{relEB}:
\begin{align}
\label{energie}
E(n,L,\gamma) = -2{n(n-1)\over L^2}\Bigg[& {c\over2}+{c^2\over12}+{n\over180}c^3 \\
&\kern-15pt+
\left({n^2\over1512}-{n\over1260}\right)c^4+...\Bigg].\nonumber
\end{align}
(For~\r{energie}, we used more terms than given above 
in $B(u)$.) Of course, this
expression agrees with~(\ref{enc}) obtained directly by expanding the~$\qa$.

\subsection{Expansion in powers of $n$}
\label{smalln}

The number of particles $n$ is \emph{a priori} an integer. However, when we
look at the small $c$ expansion~\r{Bexpn} of $B(u)$  or~\r{energie} of the
energy, we see that at  any given order in $c$ the expression is polynomial
in $n$. Therefore, one can extend the definition of the small~$c$ expansion
of~$B(u)$ or of~$E(n,L,\gamma)$ to non-integer~$n$. We can also
collect in the small $c$ expansion of $B(u)$ all the terms proportional to
$n$ and call this series $b_1(u)$.  From~(\ref{Bexpn}) we see that
\begin{align}
\label{devb1}
b_1(u) =\ & {(u^2-1)\over4}c + {(u^2 -1)^2\over96} c^2\nonumber\\
&+{(u^2-1)^2(u^2-3)\over2880}c^3 + O(c^4).
\end{align}
More generally, we can collect all the terms proportional to $n^k$ in the
small $c$ expansion and call the series $b_k(u)$. This means that we 
can write
$B(u)$ as a power series in $n$
\begin{equation}
\label{nBexpn}
B(u)=1 + n b_1(u) +n^2 b_2(u)+\dots,
\end{equation}
where all the~$b_k(u)$ are defined perturbatively in~$c$.  Conditions
(\ref{B(1)}, \ref{paritybis}) impose that all the $b_k(u)$ are even and
that $b_k(1)=0$ for all $k \geq 1$. We define $b_0(u)=1$ for consistency.
(It is easy to see in the small~$c$ expansion that if $n=0$, then $B(u)=1$.)

We are now going to describe the procedure we
used\cite{BrunetDerrida.00} to determine the whole function
$b_1(u)$ and eventually all the $b_k(u)$. If we insert
(\ref{nBexpn}) into (\ref{eqB}) we get, at first order in~$n$,
\begin{equation}
b_1(1+u)-b_1(1-u)=c\int_0^u e^{-c(v^2-uv)/2}\,dv.
\label{eqb1}
\end{equation}
It is easy to check that a solution of (\ref{eqb1}) compatible with the
conditions $b_1(1)=0$ and $b_1(u)=b_1(-u)$ is 
\begin{equation} 
b_1(u)=\sqrt c\int_0^{+\infty}\kern-1.3em d\lambda\, {\cosh\left({\lambda u\sqrt
c\over2}\right) -\cosh\left({\lambda\sqrt c\over2}\right)\over\sinh\left({\lambda\sqrt
c\over2}\right)}e^{-\lambda^2/2}.
\label{b1}
\end{equation}

There are however many other solutions of~(\ref{eqb1}), which can be
obtained by adding to~(\ref{b1}) an arbitrary function $F(u,c)$ even and
periodic in~$u$ of period~2 and vanishing at~$u=1$. If we require that each
term in the small~$c$ expansion of~$b_1(u)$ is polynomial in $u$ (as
justified in section~\ref{smallc}), we see that all the terms of the
small~$c$ expansion of~$F(u,c)$ must be identically zero.  This already
shows that~\r{b1} has the same small~$c$ expansion~(\ref{devb1}) as what
one would get by collecting all the terms proportional to~$n$ in the
small~$c$ expansion of section~\ref{smallc}.

If the solution~(\ref{b1}) of~(\ref{eqb1}) had a non-zero radius
of convergence in~$c$, it would be natural to choose this solution and 
set~$F(u,c)=0$. However it is easy to see that~(\ref{b1}) has a zero
radius of convergence in~$c$: by making the change of variable
$\lambda^2 = 2 \nu$, it is easy to see that (\ref{b1}) is the Borel sum of
a divergent series\cite{WhittakerWatson}.

Apart from being the Borel sum of its expansion in powers of~$c$, we did
not find definitive reasons why~(\ref{b1}) is the solution of (\ref{eqb1})
we should select. However, we can notice that for integer~$n$, all
the $\qa$ are real and $B(u)$ defined by~(\ref{defB}) is
analytic in $u$ and remains bounded as $|\text{Im}\ u | \to \infty$. The
solution~$b_1(u)$ given by~\r{b1} is also analytic in~$u$ and
grows as $\ln(u)$ as $|\text{Im}\ u|\to\infty$. Adding any
function~$F(u,c)$ periodic and analytic in~$u$ to~\r{b1} would
produce a much faster growth.

\medbreak
	
If we insert~(\ref{nBexpn}) into~(\ref{eqB}), we have to solve at order~$n^k$
\begin{equation}
b_k(1+u)-b_k(1-u)=\varphi_k(u),
\label{genb}
\end{equation}
where $\varphi_k(u)$ is some function odd in $u$ which can be calculated if we
know the previous orders~$b_1(u),\dots,b_{k-1}(u)$. 
\begin{align}
\label{phib}
\varphi_k(u)=
c \sum_{i=0}^{k-1} \int_0^u \kern-0.6emdv\,&e^{-c (v^2 - uv) /2}\ b_i
(1-v)\times\nonumber\\
&b_{k-i-1} (1 + u - v).
\end{align}	
We see that the difficulty of selecting a solution of a difference
equation
appears at all orders in the expansion in powers of~$n$, and we are now
going to explain the procedure we have used to select one solution.

If we write, as~$\varphi_k(u)$ is an odd function of~$u$,
\begin{equation}
\varphi_k(u)=2\int_0^{+\infty}\kern-1.3emd\lambda\,\sinh\left({\lambda u\sqrt
c\over2}\right)\ a_k(\lambda),
\label{fourier}
\end{equation}
which is equivalent, by inverting when $u$ is imaginary the Fourier
transform in~\r{fourier}, to define~$a_k(\lambda)$ by
\begin{equation}
a_k(\lambda)={1\over2i\pi}\int_0^{+\infty}\kern-1.3emdu\,\sin\left({\lambda u
\over2}\right)\ \varphi_k\left({i u\over\sqrt c}\right),
\label{aphi}
\end{equation}
then the solution for $b_k(u)$ we select is given by
\begin{equation}
b_k(u)=\int_0^{+\infty}\kern-1.3em d\lambda\, 
{\cosh{\left(\lambda u\sqrt c\over2\right)}
-\cosh{\left(\lambda\sqrt c\over2\right)}\over\sinh{\left(\lambda\sqrt
c\over2\right)}}a_k(\lambda).
\label{solgenb}
\end{equation}
Indeed, $b_k(u)$ is an even function, vanishes at~$u=1$ and one can
check using~\r{fourier} that~\r{solgenb} solves~\r{genb}.

The integrals in~(\ref{fourier}--\ref{solgenb}) are
convergent\cite{BrunetDerrida.00} and
equations~(\ref{phib}, \ref{aphi},
\ref{solgenb}) give an automatic way of calculating the $b_k(u)$ up to any
desired order.

This procedure is the direct generalisation of the choice~\r{b1} we did to
solve~\r{eqb1}. In fact, for~$k=1$, equations~(\ref{phib}, \ref{aphi}) give
(for~$\lambda\ge0$)
$a_1(\lambda)=\sqrt c \exp(-\lambda^2/2)$ and~\r{solgenb} is identical
to~\r{b1}.

As for~\r{b1}, the solution~\r{solgenb} is not the only solution
of~\r{genb}. At any order~$k$, we could add an arbitrary even periodic
function~$F(u,c)$ of period~2, the expansion of which vanishes to all order
in~$c$.  As for $b_1(u)$, we did not find an unquestionable justification
of our choice. One can notice nevertheless that~\r{solgenb} is the solution
of~\r{genb} analytic in~$u$ and with the slowest growth
when~$|\text{Im}\,u|\to\infty$.

At order $n^2$, the procedure~(\ref{phib}, \ref{aphi}) gives
\begin{align}
\label{a2}
a_2(\lambda)
=c e^{-\lambda^2/2}\Bigg[&\int_0^{\lambda}\kern-0.8em
d\mu\,e^{-\mu^2/2}{2\cosh\left({\lambda\mu\over2}\right)-2\over\tanh\left({\mu\sqrt
c\over2}\right)}\\
&+\int_\lambda^{+\infty}\kern-1.7em
d\mu\,e^{-\mu^2/2}{e^{-\lambda\mu/2}-2\over\tanh\left({\mu\sqrt
c\over2}\right)}\Bigg],\nonumber
\end{align}
with $b_2(u)$ given by~\r{solgenb}.
Writing down $b_3(u)$ or $a_3(u)$ would take here about half a column.

We can now give the first terms in the small~$n$ expansion of the energy.
Using relation~\r{relEB}, we find
\begin{align}
\label{exactE}
{L^2\over 2}E(n,L,\gamma)=\ &n\left({c\over2}+{c^2\over12}\right)\\
&-n^2
{c^{3/2}\over4}\int_0^{+\infty}\kern-1.7em 
d\lambda{\lambda^2\over\tanh\left({\lambda\sqrt c\over2}\right)}
e^{-{\lambda^2/2}}\nonumber\\
&-n^3 {c^2\over4}\int_0^{+\infty}\kern-1.7em
d\lambda{\lambda^2\over\tanh\left({\lambda\sqrt c\over2}\right)}
e^{-{\lambda^2/2}}\Bigg(\nonumber\\
&\qquad\ \int_0^{\lambda}\kern-0.8em
d\mu\,e^{-{\mu^2/2}}{2\cosh\left({\lambda\mu\over2}\right)-2\over\tanh\left({\mu\sqrt
c\over2}\right)}+\nonumber\\
&\qquad\ \int_\lambda^{+\infty}\kern-1.7em
d\mu\,e^{-{\mu^2/2}}{e^{-{\lambda\mu/2}}-2\over\tanh\left({\mu\sqrt
c\over2}\right)}\Bigg)+{n^3c^2\over6}\nonumber\\
&+O(n^4).\nonumber
\end{align}
By making the change of variable $\lambda^2=2\nu$, the terms of order~$n^2$
and~$n^3$ appear as Borel transforms of series in~$c$ with a finite radius of
convergence. We conclude that these terms have both a zero radius of
convergence in~$c$.

This small~$n$ expansion gives quickly very complicated expressions
of~$b_k(u)$. It turns out, as we shall see in the next section, that for
large~$c$, the expressions of the~$b_k(u)$ get simpler and the
energy~$E(n,L,\gamma)$ can be calculated to all orders in powers of~$n$.

\section{Expansion in powers of~$n$ in the regime~$c\to\infty$}
\label{scalingregime}

In the previous section, we have developed a procedure allowing to get the
small~$n$ expansion of the energy by solving the
problem~(\ref{eqB}--\ref{paritybis}).
Here, we show how this procedure gets greatly simplified for large~$c$.

The expansion in powers of~$n$ of the previous section  
can be summarised as follows: if we use~\r{nBexpn} and we write
\begin{equation}
a(\lambda)=na_1(\lambda)+n^2 a_2(\lambda)+\dots,
\end{equation}
the~$b_k(u)$ and~$a_k(\lambda)$ can be obtained by expanding in powers
of~$n$ the following two equations
\begin{equation}
\label{relBa}
B(u)=1+\int_0^{+\infty}\kern-1.3em d\lambda\, 
{\cosh{\left(\lambda u\sqrt c\over2\right)}
-\cosh{\left(\lambda\sqrt c\over2\right)}\over\sinh{\left(\lambda\sqrt
c\over2\right)}}a(\lambda),
\end{equation}
(this is a rewriting of~(\ref{solgenb})),
and
\begin{align}
\label{relaphi1}
&a(\lambda)={nc\over 2i\pi}\int_0^{+\infty}\kern-1.3emdu\,\sin{\left(\lambda
u\over2\right)}\times\\
&\qquad\quad\int_0^{iu\over\sqrt c} \kern-0.3em
dv\,e^{-c(v^2-{iuv/\sqrt c})/2}B(1-v)B(1+{iu\over\sqrt c}-v).\nonumber
\end{align}
(This is a rewriting of~(\ref{phib}, \ref{aphi}).)
It will be convenient in the following to replace~\r{relaphi1} by its
Fourier transform
\begin{align}
\label{relaphi}
&2\int_0^{+\infty}\kern-1.3emd\lambda\,\sinh{\left(\lambda u\sqrt
c\over2\right)}a(\lambda)=\\
&\qquad nc\int_0^u \kern-0.8em
dv\,e^{-c(v^2-uv)/2}B(1-v)B(1+u-v).\nonumber
\end{align}
(This is a rewriting of~(\ref{phib}, \ref{fourier}).)

We are going to see how one can simplify~(\ref{relBa}--\ref{relaphi})
when~$c$ is large.
First we observe that for large~$c$ and~$u$ fixed of order~1, the
expression~$b_1(u)$ takes the scaling form
\begin{equation}
b_1(1+{u\over\sqrt c})\simeq \sqrt c \int_0^{+\infty} (e^{\lambda u/2}-1)
e^{-{\lambda^2/2}}\,d\lambda.
\end{equation}
One can check from~(\ref{phib}, \ref{aphi}, \ref{solgenb}) that this
scaling form is present at any order in the small~$n$ expansion.
Indeed,~\r{relBa} becomes in the large $c$ limit
\begin{equation}
\label{relBalim}
B(1+{u\over\sqrt c})=1+\int_0^{+\infty}\kern-1.3em d\lambda\, (e^{\lambda
u/2}-1) a(\lambda),
\end{equation}
and using~\r{relaphi}, we find
\begin{align}
\label{relBB}
&2\int_0^{+\infty}\kern-1.3emd\lambda\,\sinh{\left(\lambda
u\over2\right)}\ a(\lambda)=\\
&\quad n\sqrt{c}\int_0^u\kern-1em dv\, e^{-(v^2-uv)/2}B\left(1-{v\over\sqrt c}\right)
B\left(1+{u-v\over\sqrt c}\right).\nonumber
\end{align}
It is apparent on~\r{relBalim} and~\r{relBB} that in the large $c$ limit
the function $B(1+u/\sqrt c)$ depends only on $u$ and $n\sqrt c$,
and~$a(\lambda)$ depends only on~$\lambda$ and~$n\sqrt c$.
Let us introduce the constant~$K$
\begin{equation}
K=1-\int_0^{+\infty}\kern-1.3em d\lambda\,a(\lambda).
\end{equation}
Equation~\r{relBalim} becomes
\begin{equation}
\label{relbetaa}
B(1+{u\over\sqrt c})=K+\int_0^{+\infty}\kern-1.3em d\lambda\, e^{\lambda
u/2} a(\lambda).
\end{equation}
In~\r{relBB}, if we write the integral from~0 to~$u$ as the difference
between an integral from~0 to~$+\infty$ and an integral from~$u$
to~$+\infty$, and if we change the variable in the second integral to shift
it to~0 to~$+\infty$, we obtain
\begin{align}
\label{aBB}
&2\int_0^{+\infty}\kern-1.3emd\lambda\,\sinh{\lambda
u\over2}a(\lambda)=\\
&\qquad n\sqrt{c}\int_0^{+\infty}\kern-1em dv\,
e^{-v^2/2}B(1-{v\over\sqrt c})\Bigg[e^{uv/2}B(1+{u-v\over\sqrt c})
		\nonumber\\
&\qquad\qquad\qquad-e^{-{uv/2}}B(1-{u+v\over\sqrt c})\Bigg].\nonumber
\end{align}
If we replace $B(1+(u-v)/\sqrt{c})$ and $B(1-(u+v)/\sqrt c)$ by their
expression~\r{relbetaa}, we get after some rearrangements
\begin{align}
&2\int_0^{+\infty}\kern-1.3emd\lambda\,\sinh{\lambda
u\over2}a(\lambda)=\\
&\qquad n\sqrt{c}\int_0^{+\infty}\kern-1em dv\,
e^{-v^2/2}B(1-{v\over\sqrt c})\Bigg[2K\sinh{\left(uv\over2\right)}+\nonumber\\
&\qquad\qquad\qquad\int_0^{+\infty}\kern-1.3em d\mu\,a(\mu)
e^{-{\mu v/2}}2\sinh \left(u{v+\mu\over2}\right)\Bigg].\nonumber
\end{align}
Taking the Fourier transform of this expression for imaginary~$u$, we get 
for~$\lambda\ge0$
\begin{align}
a(\lambda)=& n\sqrt{c}\int_0^{+\infty}\kern-1em dv\,
e^{-v^2/2}B(1-{v\over\sqrt c})\Bigg[K\delta(\lambda-v)+\nonumber\\
&\qquad\ \int_0^{+\infty}\kern-1.3em d\mu\,a(\mu)
e^{-{\mu v/2}}\delta(\lambda-v-\mu)\Bigg].
\end{align}
This last expression can be used to calculate~$B(1+u/\sqrt c)$
using~\r{relbetaa}:
\begin{align}
&B(1+{u\over\sqrt c})=K+\\
&\qquad n\sqrt{c}\int_0^{+\infty}\kern-1em dv\,
e^{-v^2/2}B(1-{v\over\sqrt c})\Bigg[K e^{vu/2}+\nonumber\\
&\qquad\qquad\qquad\int_0^{+\infty}\kern-1.3em d\mu\,a(\mu)
e^{-{\mu v/2}}e^{(v+\mu)u/2}\Bigg].\nonumber
\end{align}
Finally, using~\r{relbetaa}, we recognise the relation
\begin{align}
\label{betafinal}
&B(1+{u\over\sqrt c})=K+\\
&\qquad n\sqrt{c}\int_0^{+\infty}\kern-1em dv\,
e^{-v^2/2}B(1-{v\over\sqrt c})e^{vu/2}B(1+{u-v\over\sqrt
c}).\nonumber
\end{align}

We see that, in the large~$c$ limit,~(\ref{relBa}, \ref{relaphi1}) reduce
to this single equation~\r{betafinal}. We are now going to see
that~\r{betafinal} can be solved to all order in the parameter~$n\sqrt c$.
If we introduce the
function $\beta(u)$ and the parameter $\epsilon$ defined by
\begin{equation}
\beta(u)={1\over 2K\sqrt\pi} e^{-{u^2/4}}B(1+{u\over\sqrt c}),
\label{relGB}
\end{equation}
and
\begin{equation}
\epsilon=2nK\sqrt {\pi c},
\label{defepsilon}
\end{equation}
then \r{betafinal} simply becomes
\begin{equation}
\beta(u)={1\over2\sqrt\pi}e^{-{u^2/4}}
	+\epsilon\int_{0}^{+\infty}\kern-1.3emdv\, \beta(u-v) \beta(-v).
	\label{relGG}
\end{equation}

Using~(\ref{B(1)}, \ref{relEB}, \ref{relGB}), we can express the ground
state energy~$E(n,L,\gamma)$ in terms of~$\beta(u)$:
\begin{equation}
E(n,L,\gamma) = {2 \over L^2} \left[{n^3c^2\over6}+{n c^2\over12}
 - n c {\beta''(0)\over \beta(0)}\right].
\label{relEG}
\end{equation}

It is clear that relation~\r{relGG} alone determines
$\beta(u)$, at least perturbatively in $\epsilon$. So, from~\r{relEG}, we
only need to extract~$\beta(0)$ and~$\beta''(0)$ from~\r{relGG}.

It is easy to do it for the first orders in~$\epsilon$ directly from
equation~\r{relGG}. Moreover, we have found a way of calculating~$\beta(0)$
and~$\beta''(0)$, and hence the energy, to all orders in~$\epsilon$. The
details of the calculation are given in appendix~\ref{energyscaling}. The final
result can be written as
\begin{align}
\label{valE}
n\sqrt c &= {1\over2\sqrt\pi}\sum_{k=1}^{+\infty} {\epsilon^k\over k^{3/2}},\\
E(n,L,\gamma)
	&={2\over L^2}\left[{nc^2\over12}+{\sqrt c\over4\sqrt\pi}\sum_{k=1}^{+\infty}
{\epsilon^k\over k^{5/2}}\right]\label{valE2}.
\end{align}

We see that the energy is defined in an implicit way: expression~\r{valE} 
allows
to calculate $\epsilon$ as a function of $n\sqrt c$, and~\r{valE2}
gives the energy as a function of~$\epsilon$. If we substitute~$c$
using~\r{cdef}, we obtain the result announced in~\r{mainresult}.

For small~$n$, one can eliminate~$\epsilon$ from~\r{valE} and~\r{valE2}. We
get
\begin{align}
\label{Edirect}
&{L^2\over2}E(n,L,\gamma)-{nc^2\over12}={\sqrt c\over4\sqrt\pi}\Bigg(2n\sqrt{c\pi}\\
&\qquad -{\sqrt2\over8}(2n\sqrt{c\pi})^2
+\left({1\over8}-{2\sqrt3\over27}\right)(2n\sqrt{c\pi})^3
\nonumber\\
&\qquad+O\left((n \sqrt c)^4\right)\Bigg).\nonumber
\end{align}

\section{Conclusion}

In this paper, we have calculated, using the replica method, the first
cumulants~(\ref{cumulants}, \ref{exactE})  of the free energy of a directed
polymer in a random medium~(\ref{path}) for a cylinder geometry. We used
the integral equation~(\ref{eqB}) of~\cite{BrunetDerrida.00} which
together with conditions~(\ref{B(1)}, \ref{paritybis}) allowed us to expand
the moments $\langle Z^n \rangle$ of the partition function in powers
of the strength~$c$ of the disorder or in powers of the number~$n$ of
replica. All the coefficients of the small~$c$ expansion~\r{energie} are
polynomial in~$n$ allowing to define the expansions for non-integer~$n$. On
the other hand, the coefficients of the expansion~\r{exactE} in powers
of~$n$ are complicated functions of~$c$, with in general a zero radius of
convergence at~$c=0$. As already mentioned in~\cite{BrunetDerrida.00},
we think that weak disorder expansions of the moments~$\langle Z^n\rangle$ have
generically a zero radius of convergence for non-integer~$n$ when the
disorder is Gaussian; this is already the case for a single Ising spin in a
Gaussian random field\cite{BrunetDerrida.00}.

To obtain our small~$n$ expansion, we solved a difference
equation~(\ref{eqB}) which at each order in powers of~$n$ has several
solutions. We selected the particular  solution which has the slowest growth
in the imaginary~$u$ direction and has the right small~$c$ expansion,
but we could not exclude other solutions. A different approach, with a
direct calculation of the first cumulants of the free energy, and not based
on replica would therefore be very useful to test the validity of our
expressions~\r{exactE} which we have been able to derive only
perturbatively to all orders in~$c$.

Although our expansion in powers of~$n$ becomes quickly very complicated,
it simplifies when~$c$ is large and we could write in this limiting case,
all the terms of the small~$n$ expansion~(\ref{valE}, \ref{valE2}). 
The expression~(\ref{mainresult})
we obtain of the energy~$E(n,L,\gamma)$ (that is, through~\r{defE}, the
expression of~$\langle Z^n \rangle $) is given exactly by the same scaling
function as found for the ASEP. The present work therefore  gives
additional
evidence that the scaling function $G(\beta)$ given by~(\ref{largedev},
\ref{largedev2}) is characteristic of the long time behaviour of the KPZ
equation in~1+1 dimensions on a ring and that the probability distribution
of the free energy for a very long directed polymer on a ring should have
an universal shape in the range where the fluctuations per unit length of
the free energy are of order~$1/L$.  Other universal distributions for the
free energy of a directed polymer have been found recently for different
geometries\cite{PraehoferSpohn.99,Johansson.99,KimMooreBray.91,Halpin-Healy.91,KrugMeakinHalpin-Healy.92}.  Our present approach,
based on the Bethe ansatz is, at the  moment,  unable to recover these
other distributions. One can try however to extend it to open boundary
conditions (in this case too, the Bethe ansatz can be used\cite{Gaudin.83})
instead of periodic boundary conditions and see how this change of boundary
conditions affects the distribution of~$\ln Z$.  Of course, it would be
very nice to find a simpler approach which would somehow unify all these
results and allow to relate all these universal distributions corresponding
to the possible geometries, in the spirit of critical phenomena in two
dimensions where conformal invariance\cite{Cardy.84} allows to connect the
properties of different geometries.

Technically, the approach followed in the present work is simply to try to
find the~$\qa$ solution of~\r{AnsatzSol} and to calculate the
energy~\r{defener} which is a symmetric function of the roots~$\qa$, in
such a way that~$n$ becomes a continuous variable. One could do the same
in all kinds of situations. For example, in appendix~\ref{HermiteApp}, we
show how to define and calculate symmetric functions of the roots of Hermite
polynomials when the degree of the polynomial becomes non-integer.

Another interesting extension of the present work would be to consider more
general correlations of the noise~(\ref{corr}). The corresponding quantum
problem becomes then the general problem of quantum particles interacting
with an arbitrary pair potential. If the interactions are short ranged, one
expects the universality class of the KPZ equation to hold, so one could
try to repeat our expansion in powers of $c$ for a general potential
(without the use of the Bethe ansatz) simply by a standard perturbation
theory in the strength of the potential. We believe that at any order in the
strength of the potential, the ground state energy is polynomial in
$n$ allowing to define the perturbation expansion for non-integer~$n$ as we
did here.  If, with such an approach based on perturbation theory, one
could recover the scaling function $G$ of~(\ref{largedev},
\ref{largedev2}), one could try to extend the approach to higher dimension
as the relation between the directed polymer problem and the quantum
hamiltonian is valid in any dimension.

\bigbreak

{\bf Acknowledgements}
We thank Fran\c{c}ois David, Michel Gaudin, Vincent Pasquier, Herbert Spohn
and Andr\'e Voros
for interesting discussions.
\annexes

\section{Derivation of (\ref{eqB}, \ref{B(1)}, \ref{paritybis}, \ref{relEB})}
\label{derivation}

Let us first establish some useful properties of the numbers $\rho(\qa)$
defined by~(\ref{defrho}). If the $\qa$ are the $n$ roots of the
polynomial $P(X)$ 
\begin{equation}
P(X) = \prod_{\qa} (X - \qa),
\label{defpoly1}
\end{equation}
it is easy to see that the $\rho(\qa)$ defined in (\ref{defrho})
satisfy
\begin{equation}
{P(X+c)\over P(X)}=1+c\sum_{\qa} {\rho(\qa)\over X-\qa}.
\label{simpleelement}
\end{equation}
(The two sides have the same poles with the same residues and coincide 
at $X \to \infty$.) Expanding the right hand side
of~\r{simpleelement} for large $X$, we get
\begin{align}
{P(X+c) \over P(X)} =\ & 1+ 
c\sum_{\qa}{\rho(\qa)\over X}\left(1+{\qa\over X}+{\qa^2\over
X^2}\right) \nonumber\\&+ O \left( 1 \over X^4 \right).
\label{rap1}
\end{align}
On the other hand, using (\ref{defener}, \ref{defpoly1}) and the
 symmetry $\lbrace\qa\rbrace = \lbrace - \qa\rbrace$ we have
\begin{align}
P(X)= & X^n + {L^2 \over 4} E(n,L,\gamma) X^{n-2} \nonumber\\
	&+ O(X^{n-4}),
\end{align}
so that 
\begin{align}
\label{rap2}
&{P(X+c) \over P(X)} = 1 + {nc\over X} + {c^2\bin{n}{2}\over
X^2}\\
&\qquad+{c^3\bin{n}{3}-c E(n,L,\gamma) L^2 / 2 \over X^3} + O \left( 1 \over
X^4 \right).\nonumber
\end{align}
Comparing (\ref{rap1}) and (\ref{rap2}), we get the relations
\begin{align}
\label{prop1}
\sum_{\qa} \rho(\qa)&=n,\\
\sum_{\qa} \qa\rho(\qa)&=c\bin{n}{2}, \label{prop3}\\
\sum_{\qa} \qa^2\rho(\qa)&=c^2\bin{n}{3}- {E(n,L,\gamma) L
^2 \over 2}. \label{prop4}
\end{align}
Moreover, by letting $X=\pm\qb-c$ in~\r{simpleelement} 
 we get for any $\qb$ root of $P(X)$
\begin{equation}
\label{prop2}
{1\over c}=
\sum_{\qa}{\rho(\qa)\over\qa-\qb+c}
=\sum_{\qa}{\rho(\qa)\over\qa+\qb+c}.
\end{equation}
Lastly using the symmetry $\lbrace\qa\rbrace = \lbrace - \qa\rbrace$ and
the definition (\ref{defrho}), the Bethe ansatz equations
(\ref{AnsatzSol}) reduce to 
\begin{equation}
e^{\qa} \rho(-\qa)-e^{-\qa}\rho(\qa)=0.
\label{eqrho}
\end{equation}

From the definition (\ref{defB}) of $B(u)$ and the properties
(\ref{prop1}--\ref{eqrho}), it is straightforward to establish
(\ref{eqB}--\ref{relEB}): the integral equation (\ref{eqB}) is a direct
consequence of (\ref{defB}) and (\ref{prop2}). Properties
(\ref{B(1)}, \ref{paritybis}) follow from (\ref{defB}, \ref{prop1}) and
(\ref{defB}, \ref{eqrho}) respectively. Lastly (\ref{relEB}) is a
consequence of (\ref{defB}, \ref{prop1}--\ref{prop4}).

\section{The energy in the scaling regime}
\label{energyscaling}

In this appendix, we show how to calculate the energy from the integral
equation~\r{relGG}. This equation is of the form
\begin{equation}
\beta(u)=H(u) +\epsilon\int_{0}^{+\infty}\kern-1.3em dv\, \beta(u-v) \beta(-v),
\label{relGHGG}
\end{equation}
where, in our case, $H(u)$ is given by
\begin{equation}
H(u)={1\over2\sqrt\pi} e^{-{u^2\over4}}.
\label{defH}
\end{equation}
We are going to do our calculations for an arbitrary function~$H(u)$,
even in~$u$ and decreasing fast enough (to make all the integrals converge)
when~$|u|\to\infty$.

To find the energy, we see from~\r{relEG}, that we have to calculate
from~\r{relGHGG} the quantities~$\beta(0)$
and~$\beta''(0)$ as functions of~$\epsilon$.  We first show
that~\r{relGHGG} is equivalent to
\begin{equation}
\beta(u)=H(u)+\epsilon\int_0^{+\infty}\kern-1.3em dv\, H(u-v)\beta(v),
\label{relGH}
\end{equation}
as long as $H(u)$ is even and decreases fast enough. Then, we will
introduce a new function~$\betas(u)$ which is easy to calculate, and relate
the derivatives of~$\beta(u)$ and~$\betas(u)$ at~$u=0$.

\subsection{Equivalence between~\r{relGHGG} and~\r{relGH}}

The solution of~\r{relGH} can be written as
\begin{equation}
\label{expnbeta}
\beta(u)=\beta_0(u)+\epsilon \beta_1(u)+\epsilon^2\beta_2(u)+\dots,
\end{equation}
where
\begin{align}
\label{Gexpn}
\beta_0(u) &= H(u),\\
\beta_1(u) &= \int_0^{+\infty} H(u-v_1) H(v_1) \, dv_1,\nonumber\\
\beta_2(u) &= \iint_0^{+\infty} H(u-v_1) H(v_1-v_2) H(v_2)\,dv_1dv_2,\nonumber\\
\dots\nonumber\\
\beta_k(u) &= \int\kern-0.5em\cdots\kern-0.5em\int_0^{+\infty} H(u-v_1)
H(v_1-v_2)\dots\nonumber\\
&\hphantom{\int\kern-0.5em\cdots\kern-0.5em\int_0^{+\infty}}\quad
\dots H(v_{k-1}-v_k)H(v_k)\,dv_1\dots
dv_k.\nonumber
\end{align}
For a given~$k>0$, the integration range of~$\beta_k(u)$
can be divided into~$k$ parts: the region where~$v_1$
has the lowest value of all the~$\lbrace v_i\rbrace$, the region
where~$v_2$ has the lowest value, \ldots, the region where~$v_k$ has the
lowest value. Let us consider, for some~$j$ such that $1\le j\le k$,
the region where $v_j$ has the lowest value. All the other integrals then run
from $v_j$ to $+\infty$. If we translate those to integrals running from
$0$ to $+\infty$ by changing $v_i$ into $v_i+v_j$, we get:
\begin{align}
\label{kparts}
\int_0^{+\infty}\kern-1.7emdv_j\, 
	\int_0^{+\infty}\kern-1.7em dv_1\dots dv_{j-1}\,
			& H(u-v_1-v_j)\\
			&H(v_1-v_2)\dots H(v_{j-1}) \times\nonumber\\
	\int_0^{+\infty}\kern-1.7em dv_{j+1}\dots dv_k\,
			& H(-v_{j+1})\nonumber\\
			&H(v_{j+1}-v_{j+2})\dots H(v_k+v_j)\nonumber
\end{align}
Using the fact that $H(u)=H(-u)$, we see that~\r{kparts}
is equal to
\begin{equation}
\int_0^{+\infty}\kern-1.7em dv_j\, \beta_{j-1}(u-v_j) \beta_{k-j}(-v_j).
\end{equation}
By summing over~$j$, we therefore have
\begin{equation}
\beta_k(u)=\int_0^{+\infty}\kern-1.7em dv\, \sum_{j=1}^{k} \beta_{j-1}(u-v)
\beta_{k-j}(-v).
\end{equation}
Finally, if we multiply by $\epsilon^k$ and if we sum over $k$ 
all these terms (keeping apart the term for
$k=0$), we obtain equation~\r{relGHGG}.

The equations~\r{relGHGG} and~\r{relGH} are thus equivalent
and~(\ref{expnbeta}, \ref{Gexpn}) give the solution of~\r{relGHGG} to any
order in~$\epsilon$.

\subsection{Calculation of the derivatives of~$\beta(u)$}

If we look at the expression~\r{Gexpn} of $\beta(u)$ in powers of
$\epsilon$, the calculation of~$\beta(0)$ and~$\beta''(0)$ looks simple,
especially when $H(u)$
is given by~\r{defH}. However, when we try to
actually do the calculation, the expressions become quickly complicated
with error-functions, primitives of error-functions, etc. It
would be much easier if the integrals in~\r{Gexpn} were running from
$-\infty$ to $+\infty$ instead of~$0$ to $+\infty$.
This is why we introduce the even function
\begin{equation}
\betas(u)=\betas_0(u)+\epsilon\betas_1(u)+\epsilon^2\betas_2(u)+\dots,
\end{equation}
where, for~$k>0$,
\begin{equation}
\betas_k(u)= {1\over k+1}\int\kern-0.5em\cdots\kern-0.5em\int_{-\infty}^{+\infty}
\kern-1.7em H(u-v_1)\dots H(v_k)\,dv_1\dots dv_k,
\label{defGb}
\end{equation}
and $\betas_0(u)=H(u)$. One can see easily that
\begin{equation}
\label{relbsH}
\betas(u)=
{-1\over2\pi\epsilon}\int_{-\infty}^{+\infty}\kern-1.75em dq\,
e^{-iqu}\ln(1-\epsilon \Ht(q)),
\end{equation}
where we have defined
\begin{equation}
\label{defHt}
\Ht(q) = \int_{-\infty}^{+\infty}\kern-1.75em du\,e^{iqu}H(u).
\end{equation}

The Wiener-Hopf technique\cite{MMP} allows to relate~$\beta(u)$
and~$\betas(u)$. More specifically, we are going to show that for
any~$X>0$,
\begin{equation}
\label{relbbs}
\epsilon\int_0^{+\infty}\kern-1.75em du\,e^{-uX}\betas(u)=
\ln\left(1+\epsilon\int_0^{+\infty}\kern-1.75em du\,e^{-uX}\beta(u)
\right).
\end{equation}
This relation allows to relate the derivatives
of~$\beta(u)$ and~$\betas(u)$ at~$u=0$: indeed, if~$X$ is large
in~\r{relbbs}, we get
\begin{equation}
\int_0^{+\infty}\kern-1.75em du\,e^{-uX}\beta(u)
={\beta(0)\over X}+{\beta'(0)\over X^2}+{\beta^{''}(0)\over X^3}+\dots,
\end{equation}
and a similar expression for~$\betas(u)$. Comparing
both sides of~\r{relbbs} gives
\begin{align}
\label{relGbG}
\beta(0) =& \betas(0),\\
\beta'(0) =& {\epsilon\over2} \beta(0)^2,\nonumber\\
\beta''(0) =& {\betas}''(0) +{\epsilon^2\over6} \beta(0)^3.\nonumber
\end{align}
(We have used the fact that~${\betas}'(0)=0$ because~$\betas(u)$ 
is an even function.)

\bigbreak

In order to prove~\r{relbbs}, 
the first thing to note is that, as~$H(u)$ decreases fast
when~$u\to\pm\infty$, then also does~$\beta(u)$. This allows to define the
two ``partial'' Fourier transforms
\begin{align}
\betat_{+}(q)&=\int_{0}^{+\infty}\kern-1.75em du\,e^{iqu}\beta(u),\\
\betat_{-}(q)&=\int_{-\infty}^{0}\kern-1.35em du\,e^{iqu}\beta(u).
\end{align}
It is easy to see that~$\betat_{+}(q)$ is analytic in
the upper half-plane ($\text{Im}\,q\ge0$). Moreover,
in this half-plane, $\betat_{+}(q)$ is bounded and vanishes
when~$|q|\to\infty$. Conversely,~$\betat_{-}(q)$ is analytic, bounded and
decreases to~0 at infinity when~$\text{Im}\,q\le0$.

The function~$\beta(u)$ can be written in terms of~$\betat_{+}(q)$
and~$\betat_{-}(q)$:
\begin{equation}
\beta(u) ={1\over2\pi}\int_{-\infty}^{+\infty}\kern-1.75em
dq\,e^{-iqu}(\betat_{+}(q)+\betat_{-}(q)),
\end{equation}
which allows to express the right-hand side of~\r{relbbs}
when~$X$ is positive:
\begin{align}
\label{lapb}
\int_0^{+\infty}\kern-1.75em du\,e^{-uX}\beta(u)
=\ &{1\over2\pi}\int_{-\infty}^{+\infty}\kern-1.75em dq\ {\betat_{+}(q)\over
X+iq}\\
&+{1\over2\pi}\int_{-\infty}^{+\infty}\kern-1.75em dq\ {\betat_{-}(q)\over
X+iq}.\nonumber
\end{align}
We calculate the two integrals in the right hand side of~\r{lapb} by
the residue theorem. As $\betat_{+}(q)$ is analytic and decreases at
infinity in the upper half-plane, the first integral can be written
when~$X>0$ as a
contour integral around the upper half-plane. The only contribution to the
first integral comes, using Cauchy's theorem, from the pole~$q=iX$. One can
also check that the
second integral vanishes (using a contour around the lower half-plane and
the fact that~$\betat_{-}(q)$ has no pole). Therefore,~\r{lapb} gives
\begin{equation}
\int_0^{+\infty}\kern-1.75em du\,e^{-uX}\beta(u) =\betat_{+}(iX).
\label{bbslhs}
\end{equation}

Now, if we multiply~\r{relGH} by~$\exp(iqu)$ and if we integrate over~$u$, we
easily get for any real~$q$
\begin{equation}
\betat_{+}(q)+\betat_{-}(q)=\Ht(q)+\epsilon \Ht(q) \betat_{+}(q).
\end{equation}
This relation between~$\Ht(q)$, $\betat_{-}(q)$ and~$\betat_{+}(q)$,
together with~\r{relbsH} gives
\begin{align}
\betas(u)={1\over2\pi\epsilon}\int_{-\infty}^{+\infty}\kern-1.75em dq\,
e^{-iqu}\Big(&\ln(1+\epsilon\betat_{+}(q))\\
	     &-\ln(1-\epsilon\betat_{-}(q)\Big).\nonumber
\end{align}
Using again that, in the upper half-plane,~$\betat_{+}(u)$ is analytic and
vanishes at infinity, we see that, for  a \emph{small
enough}~$\epsilon$, the quantity~$\ln(1+\epsilon\betat_{+}(q))$ is also
analytic and decreases to~0 at infinity when~$\text{Im}\,q\ge0$. 
Similarly, $\ln(1-\epsilon\betat_{-}(q))$ has the same properties
for~$\text{Im}\,q\le0$. This allows to calculate the left hand side
of~\r{relbbs} as we did for the right hand side. We find
\begin{equation}
\int_{0}^{+\infty}\kern-1.75em du
e^{-uX}\betas(u)={1\over\epsilon}\ln(1+\epsilon \betat_{+}(iX)).
\label{bbsrhs}
\end{equation}

Comparing~(\ref{bbslhs}) and~\r{bbsrhs} completes the proof of~\r{relbbs}.

\bigbreak

We can now give an expression of the energy. If we use the
definition~\r{defH} of~$H(u)$ in~(\ref{relbsH}, \ref{defHt}), we find
\begin{equation}
\betas(u)={1\over2\sqrt\pi} \sum_{k=0}^{+\infty}
{\epsilon^k\over(k+1)^{3/2}}e^{-{u^2/[4(k+1)]}}.
\end{equation}
This gives
\begin{align}
\betas(0)=&{1\over2\sqrt\pi} \sum_{k=0}^{+\infty}{\epsilon^k\over(k+1)^{3/2}},
\label{valGb}\\
{\betas}''(0)=&-{1\over4\sqrt\pi}\sum_{k=0}^{+\infty}{\epsilon^k\over(k+1)^{5/2}},
\label{valGb2}
\end{align}
and, together with~\r{relGbG}, these equations allows to give an expression
of~$\beta(0)$ and~${\beta''}(0)$.

From~(\ref{B(1)}, \ref{relGB}, \ref{defepsilon}), we see that
\begin{equation}
\epsilon \beta(0)=n\sqrt c.
\end{equation}
Then, using~\r{relGbG} we get
\begin{align}
 \epsilon\betas(0)=&n\sqrt c,\\
{\beta''(0)\over \beta(0)}=&{\epsilon\over n\sqrt c} {\betas}''(0)+{n^2 c\over6}\nonumber.
\end{align}
The energy is given by~\r{relEG}. We get:
\begin{equation}
E(n,L,\gamma) = {2\over L^2}\left[{nc^2\over12}-\epsilon\sqrt c
{\betas}''(0)\right].
\end{equation}
And, finally, using relation~(\ref{valGb}, \ref{valGb2}), we 
obtain~(\ref{valE}, \ref{valE2}).

\section{Hermite polynomials with a non-integer number of roots}
\label{HermiteApp}

What we try to do in this whole paper is essentially  to calculate
$\sum_\alpha q_\alpha^2$ (the energy) where~$\lbrace\qa\rbrace$ is
solution of~(\ref{AnsatzSol}), in such a way that~$n$ appears as a
continuous
parameter. This allows us to obtain expressions of the energy for
non-integer~$n$.

One can use the same procedure in other kinds of situations. A simple example
which illustrates our calculations is the case of the zeroes of Hermite
polynomials.

The $n$-th Hermite polynomial $H_n(X)$ is the solution polynomial in~$X$
with leading coefficient~1 of the differential equation\cite{HMF}
\begin{equation}
{1\over2}H_n''(X)-X H_n'(X)+n H_n(X) =0.
\label{Hermite}
\end{equation}
The polynomial $H_n(x)$ is of degree $n$ and has the symmetry~$H_n(X)=
(-)^n H_n(-X)$. For example, we have~ $H_4(X)=X^4-3X^2+{3\over4}$.
The $n$ roots $\lbrace \ha\rbrace$ ($1\le\alpha\le n$) of $H(X)$ are real 
and distinct\cite{Szego.39}.

By deriving~\r{Hermite} $p$~times with respect to $X$, we see that, for
all $p$, 
\begin{equation}
\label{Hermiterecur}
X H_n^{(p+1)}(X) = {1\over2} H_n^{(p+2)}(X)+(n-p) H_n^{(p)}(X).
\end{equation}
This shows that the $(n-p)$-th Hermite polynomial is, up to a constant
factor, equal to the $p$-th derivative of $H_n(X)$. (This property will be used
a lot in appendix~\ref{appDirect}).

Equation~\r{Hermite} can be used directly to calculate the
first coefficients of $H_n(X)$
\begin{equation}
\label{Hermiteexpn}
H_n(X)=X^n-{1\over2}\bin{n}{2}X^{n-2}+{3\over4}\bin{n}{4}X^{n-4}+\dots
\end{equation}
Using~\r{Hermiteexpn}, the symmetry of~$H(X)$ and the large~$X$ expansion
\begin{equation}
{H_n'(X) \over H_n(X)} = \sum_{p\geq 0} {1 \over X^{p+1}}
 \left( \sum_\alpha \ha^p \right),
\end{equation}
we can calculate the moments of the roots $\lbrace\ha\rbrace$
of~$H(X)$:
\begin{align}
\label{hermitemoments1}
\sum_\alpha \ha^2 &= {n(n-1)\over2}, \\
\label{hermitemoments2}
\sum_\alpha \ha^4 &= {n(n-1)\over4}(2n-3),
\end{align}
and so on. 
These moments are a priori defined only for integer $n$ but
as the expressions are polynomial in $n$, one can obviously extend
their definition to non-integer $n$ (similarly to what we do in the 
 small $c$ expansion of~$B(u)$ in section~\r{smalln}).

To generate all the moments of the roots $\ha$, it is convenient to consider the generating function
\begin{equation}
\label{Qudef}
Q(u)=\sum_{\ha} e^{\ha u},
\end{equation}
which is reminiscent of the quantity~$\beta(u)$ defined in our
quantum problem. (Using~\r{defB} and~\r{relGB} we can check that
$\beta(u)\propto\exp(u\sqrt c/2)\sum \rho(\qa)\exp(\qa u/\sqrt c)$.)

The function $Q(u)$ is hard to calculate 
for general $n$ but we can expand it in powers of
$n$. This can be done by considering
\begin{equation}
\label{relpsiQ}
\Psi(X)={H_n'(X)\over H_n(X)} = \int_0^{+\infty}\kern-1.3em du\, Q(u) e^{-u X},
\end{equation}
which is defined only for $X$ positive and large enough to make the
integral converges.  This function $\Psi(X)$ is solution of a differential
equation which follows from~\r{Hermite}:
\begin{equation}
{1\over2}\Psi'(X)+{1\over2}\Psi(X)^2- X \Psi(X)+n=0.
\label{Psieq}
\end{equation}
To obtain an expansion in powers of $n$, we write
\begin{equation}
\label{defpsi2}
\Psi(X)=n\Psi_1(X)+n^2\Psi_2(X)+\dots
\end{equation}
Thus $\Psi_1(X)$ satisfies
\begin{equation}
{1\over2}\Psi_1'(X) - X \Psi_1(X)+1=0.
\end{equation}
This differential equation can easily be solved, and the integration
constant can be fixed using the requirement (\ref{relpsiQ}) that, for
large~$X$, $\Psi(X) \simeq n/X$ 
\begin{equation}
\Psi_1(X)=\int_0^{+\infty}\kern-1.7em du\,e^{-uX-{u^2/4}}.
\end{equation}
Then order $n^2$ of (\ref{Psieq}) gives
\begin{equation}
{1\over2}\Psi_2'(X)-X\Psi_2(X)+{1\over2}\Psi_1(X)^2=0,
\end{equation}
 the solution of which can be written as
\begin{equation}
\Psi_2(X)=2\int_0^{+\infty}\kern-1.7em
du\,e^{-uX-{u^2/4}}\int_0^{+\infty}\kern-1.1em dt\, {\cosh
{\left(ut\over\sqrt2\right)}-1\over t}e^{-t^2}.
\end{equation}
The procedure can be iterated to any order in $n$ (of course expressions 
become more and more complicated). Using~\r{relpsiQ} and the expressions of
$\Psi_1(X)$ and $\Psi_2(X)$ we can give an expression of $Q(u)$:
\begin{align}
\label{Qures}
Q(u)=\ & n e^{-{u^2/4}}+\\
&2 n^2 e^{-{u^2/4}}\int_0^{+\infty}\kern-1.1em
dt\, {\cosh {\left(ut\over\sqrt2\right)}-1\over t}e^{-t^2}+O(n^3).\nonumber
\end{align}
Expanding this expression in powers of $u$, one calculate from this
expression and from (\ref{Qudef}) the terms linear and quadratic in $n$ of
all the moments of the $\ha$. (The results agree for the second and the
fourth moments with (\ref{hermitemoments1}, \ref{hermitemoments2}).)

We noticed that for small $n$, the expression (\ref{Qures}) corresponds to
$n$ roots $\ha$ distributed along the imaginary axis with a Gaussian
distribution. We do not know whether this is general and whether there exists, for
general non-integer $n$, a distribution of the roots $\ha$ in the complex
plane plane which gives all moments calculated as in
(\ref{hermitemoments1}, \ref{hermitemoments2}).

It is interesting to notice the similarity between $Q(u)$ and $\beta(u)$
defined in section~\ref{scalingregime}.

\section{The expansion in powers of $c$ using Hermite polynomials}
\label{appDirect}

In this appendix we show how to expand the solution~$\{ \qa \}$
of~(\ref{AnsatzSol}) in powers of $c$ for integer $n$.  One can see
from~(\ref{AnsatzSol}) that the roots $\qa$ scale for small $c$ like $\sqrt
c$.  It is thus convenient to rescale the polynomial $P(X)$ defined in
(\ref{defpoly}) and the $\qa$ in the following way:
\begin{align}
\label{defR}
\qa &= \ra\sqrt c, \\
P(X\sqrt c) &= c^{n/2} R(X). \nonumber
\end{align}
($\lbrace\ra\rbrace$ are thus the roots of $R(X)$.)
With these new variables, equation~\r{mainqa} becomes
\begin{equation}
e^{\ra\sqrt c}R(\ra-\sqrt c)+e^{-\ra\sqrt c} R(\ra+\sqrt c)=0.
\end{equation}
As the roots $\ra$ of $R(X)$ are all distinct, this equation is
equivalent to
\begin{align}
\label{mainRf}
&e^{X\sqrt c}R(X-\sqrt c)+e^{-X\sqrt c} R(X+\sqrt c)=\\
&\qquad2(\cosh X\sqrt c+f(X))R(X),\nonumber
\end{align}
where $f(X)$ is \emph{analytic} (this follows from the fact that as $R(X)$
is polynomial, $f(X)$ defined by~\r{mainRf} is obviously meromorphic; moreover
as the left hand side of~\r{mainRf} vanishes at all the roots of $R(X)$,
$f(X)$ has no pole.) We are now going to solve~(\ref{mainRf}) as a power
series in $c$ (\textit{i.e.} find both $f(X)$ and $R(X)$ as power series in
$c$).

\subsection{Expansion of the polynomial~$R(X)$}
We only have the single equation~\r{mainRf} to obtain two quantities 
($R(X)$ and $f(X)$); however, using the fact that $f(X)$ has no pole and $R(X)$
is a polynomial, 
 both quantities can be determined in a small $c$
expansion. Let us write
\begin{align}
R(X) &= R_0(X) + c R_1(X) + c^2 R_2(X)+\dots,\\
f(X) &= \hphantom{R_0(X) + {}} c f_1(X) + c^2 f_2(X)+\dots,\nonumber
\end{align}
where the $f_i(X)$ have no pole, $R_0(X)$ is a polynomial of degree $n$ 
(the term of highest degree in $R_0(X)$ is $X^n$)
and all the $R_i(X)$ (for $i\ge1$) are polynomials of degree less
than $n$.
At first order in $c$, we find that~\r{mainRf} gives:
\begin{equation}
{1\over2}R_0''-X R_0' = f_1 R_0.
\label{eqR0}
\end{equation}
As $f_1(X)$ has no pole, it must be a polynomial.
Because $R_0(X)$ is of degree $n$, we see by looking at both
sides of (\ref{eqR0}) that, necessarily, $f_1(X)=-n$. We recognise
then the differential equation~\r{Hermite} that defines Hermite polynomials.
Therefore 
\begin{align}
f_1(X)=&-n,\\
R_0(X)=&H(X).\nonumber
\end{align}
We recover that way that the $\ra$ are the zeroes of the~$n$-th
Hermite polynomial when $c$ is very small\cite{Gaudin.83}.

At next order in $c$, equation~\r{mainRf} gives
\begin{align}
\label{Hermiteorder2}
&{1\over2}R_1''-X R_1'+n R_1-f_2 H=\\
&\qquad {X^3\over6}H'-{X^2\over4}H''+{X\over6}H^{(3)}
-{1\over24}H^{(4)}.\nonumber
\end{align}
As $R_1$ and $H$ are polynomials, 
\r{Hermiteorder2}
 tells us that $f_2 H$ is a polynomial too.
We also know that $f_2(X)$ has no pole, thus it must be a polynomial.
$R_1(X)$ is of degree strictly less than $n$, so the expression
$R_1''/2-XR_1+nR_1$ is of degree strictly less than $n$. As $H$ is of
degree $n$, we recognise in~\r{Hermiteorder2} an euclidian division of
polynomials: $-f_2(X)$ is the quotient of the right hand side of
equation~\r{Hermiteorder2} divided by $H(X)$, and the terms involving
$R_1(X)$ form the remainder of this division. This ensures that there is
only one possible function $f_2(X)$ which verifies~\r{Hermiteorder2}.

In practise, to perform this euclidian division we can use
the property~\r{Hermiterecur} of the Hermite polynomials 
 as many times as needed in the right hand side
of~\r{Hermiteorder2}: for instance, we transform the term
$X^3 H'/6$ into $n X^2 H/6 + X^2 H''/12$. We cannot change $X^2 H$ anymore, 
but we can apply~\r{Hermiterecur} to the term $X^2 H''$.
When no more transformation is possible, we are left with:
\begin{align}
&{X^3\over6}H'-{X^2\over4}H''+{X\over6}H^{(3)}-{1\over24}H^{(4)}=\\
&\qquad\left({n\over6}X^2-{n(n-1)\over6}\right)H-{1\over12}H''.\nonumber
\end{align}
The Euclidian division is then easy to perform
\begin{align}
&f_2(X)=-{n\over6}X^2+{n(n-1)\over6},\\
&{1\over2}R_1''-X R_1'+n R_1=-{1\over12}H''.\nonumber
\end{align}
Using again~\r{Hermiterecur}, the differential equation on $R_1$ can be
solved; we find
\begin{equation}
R_1(X)=-{1\over24}H''(X).
\end{equation}
As $R_1(X)$ is simply a derivative of $H(X)$, and as $f_1(X)$ is a known
polynomial of $X$, we see that at the next order
in $c$ we will have to solve an equation of the form
\begin{equation}
{1\over2}R_2''-X R_2 +n R_2 -f_3 H=\sum X^j H^{(k)}.
\end{equation}
Using many times equation~\r{Hermiterecur} the right hand side can be
written in a ``canonical form'':
\begin{equation}
\sum X^j H^{(k)}=\sum X^j H + \sum H^{(k)},
\end{equation}
which allows to write $f_3$ as a polynomial in $X$ and $R_2$ as a sum of
derivatives of $H(X)$.  It is easy to see recursively that at any order
$c^k$ in the expansion we can repeat this procedure to calculate $f_k(X)$
and $R_{k-1}(X)$.  As a result we see that $f_k$ is a polynomial in $X$ and
that $R_{k-1}$ can be written as a sum of derivatives of $H(X)$.

It is worth noting that at each order the variable $n$ comes from the
previous orders
and from transformations of the kind~$X H'(X) \to {1\over2}H''(X)+nH(X)$.
Because those are the two only mechanisms by which $n$ appears, it is easy to
see that at each order the coefficients of the sum of derivatives of
$H(X)$ that constitutes $R_{k-1}(X)$ are all \emph{polynomials in $n$}.

A computer can easily do this tedious but straightforward task to any
desired order. Up to $c^3$, we find:
\begin{align}
\label{RinH}
R=&H-{c\over24}H''-c^2\left({n\over360}H''-{7\over5760}H^{(4)}\right)+
\nonumber\\
&c^3\Bigg(\left({n\over2520}-{n^2\over3024}\right)H''+{11n\over60480}H^{(4)}
\nonumber\\
&\qquad\qquad-{31\over967680}H^{(6)}\Bigg)+O(c^4).
\end{align}

\subsection{Expansion of the roots $\ra$ of $R(X)$}

As seen in~\r{RinH}, the polynomial $R(X)$ is to leading order in $c$ given
by $H(X)$. It is thus natural to write the roots $\ra$ of $R(X)$ as
\begin{equation}
\ra=\ha+c x_\alpha +O(c^2).
\label{raexp}
\end{equation}
($\lbrace\ha\rbrace$ are the roots of $H$).
Inserting~\r{raexp} into~\r{RinH}, we find, at first order in~$c$,
\begin{equation}
x_\alpha H'(\ha)-{1\over24}H''(\ha)=0.
\end{equation}
Using the definition~\r{Hermite} of Hermite polynomials, we have
$H''(\ha)=2\ha H'(\ha)$. This gives in turn~$x_\alpha={1\over12}\ha$.
Repeating this procedure to any order in~$c$, we generate terms of the
form $\ha^j H^{(k)}(\ha)$ which can be reduced to terms of the form $\ha^l
H'(\ha)$ by using~\r{Hermiterecur} as many times as necessary. It is then
possible to divide the expression by $H'(\ha)$ and we are left with an
equation giving each new term in the expansion of $\ra$ as a
\emph{polynomial in $\ha$}. Again, this can be programmed, and we get, up
to the order $c^2$:
\begin{align}
\label{rares}
\ra={\qa\over\sqrt c}=\ha+{c\over12}\ha+
c^2\Bigg(&\left({n\over120}-{11\over1440}\right)\ha\nonumber\\
&-{1\over360}\ha^3\Bigg)+O(c^3).
\end{align}
 Using~\r{defR} and~\r{defener}, this leads to
\begin{align}
&{2\over L^2}E(n,L,\gamma)=-c\sum\ha^2-{c^2\over6}\sum\ha^2 -\\
&\qquad{c^3\over360}\left((6n-3)\sum\ha^2-2\sum\ha^4\right)+O(c^4).\nonumber
\end{align}
which coincides with~\r{energie} when one uses the
properties~(\ref{hermitemoments1}, \ref{hermitemoments2}) of the
roots~$\ha$ of the Hermite polynomials.


\begin{thebibliography}{10}

\bibitem{KardarZhang.87}
M. Kardar and Y.-C. Zhang, ``Scaling of Directed Polymers in Random Media,''
  Phys. Rev. Lett. {\bf 58,} 2087--2090 (1987).

\bibitem{Zhang.87}
Y.-C. Zhang, ``Ground-State Instability of a Random System,'' Phys. Rev. Lett.
  {\bf 59,} 2125--2128 (1987).

\bibitem{Halpin-HealyZhang.95}
T. Halpin-Healy and Y.-C. Zhang, ``Kinetic Roughening Phenomena, Stochastic
  Growth, Directed Polymers and all That,'' Physics Reports {\bf 254,} 215--414
  (1995).

\bibitem{DerridaSpohn.88}
B. Derrida and H. Spohn, ``Polymers on Disordered Trees, Spin Glasses and
  Traveling Waves,'' J. Stat. Phys. {\bf 51,} 817--840 (1988).

\bibitem{Derrida.91}
B. Derrida, ``Mean Field Theory of Directed Polymers in a Random Medium,''
  Physica Scripta {\bf T38,} 6--12 (1991).

\bibitem{MezardParisiVirasoro.87}
M. M{\'e}zard, G. Parisi, and M. Virasoro, {\em Spin Glass Theory and Beyond}
  (World Scientific, 1987).

\bibitem{ImbrieSpencer.88}
J.~Z. Imbrie and T. Spencer, ``Diffusion of Directed Polymers in a Random
  Environment,'' J. Stat. Phys. {\bf 52,} 609--626 (1988).

\bibitem{KardarParisiZhang.86}
M. Kardar, G. Parisi, and Y.-C. Zhang, ``Dynamic Scaling of Growing
  Interfaces,'' Phys. Rev. Lett. {\bf 56,} 889--892 (1986).

\bibitem{Krug.97}
J. Krug, ``Origins of Scale Invariance in Growth Processes,'' Adv. Phys. {\bf
  46,} 139--282 (1997).

\bibitem{Kardar.87}
M. Kardar, ``Replica {B}ethe {A}nsatz Studies of Two-Dimensional Interfaces
  with Quenched Random Impurities,'' Nuclear Physics B {\bf 290 [FS20],}
  582--602 (1987).

\bibitem{DerridaLebowitz.98}
B. Derrida and J.~L. Lebowitz, ``Exact Large Deviation Function in the
  Asymmetric Exclusion Process,'' Phys. Rev. Lett. {\bf 80,} 209--213 (1998).

\bibitem{DerridaAppert.99}
B. Derrida and C. Appert, ``Universal Large-Deviation Function of the
  {K}ardar-{P}arisi-{Z}hang Equation in One Dimension,'' J. Stat. Phys. {\bf
  94,} 1--30 (1999).

\bibitem{Kim.95}
D. Kim, ``{B}ethe Ansatz Solution for Crossover Scaling Functions of the
  Asymmetric {XXZ} Chain and the {K}ardar-{P}arisi-{Z}hang-Type Growth Model,''
  Phys. Rev. E {\bf 52,} 3512--3524 (1995).

\bibitem{LeeKim.99}
D.-S. Lee and D. Kim, ``Large Deviation Function of the Partially Asymmetric
  Exclusion Process,'' Phys. Rev. E {\bf 59,} 6476--6482 (1999).

\bibitem{LebowitzSpohn.99}
J.~L. Lebowitz and H. Spohn, ``A {G}allavotti-{C}ohen-Type Symmetry in the
  Large Deviation Functional for Stochastic Dynamics,'' J. Stat. Phys. {\bf
  95,} 333--365 (1999).

\bibitem{Appert.00}
C. Appert, ``Large Deviation Function for the {E}den Model and Universality
  within the One-Dimensional {K}ardar-{P}arisi {Z}hang class,'' Phys. Rev. E
  {\bf 61,} 2092--2094 (2000).

\bibitem{BrunetDerrida.00}
{\'E}. Brunet and B. Derrida, ``Ground state energy of a non-integer number of
  particles with $\delta$ attractive interactions,'' Physica A {\bf 279,}
  398--407 (2000).

\bibitem{LiebLiniger.63}
E.~H. Lieb and W. Liniger, ``Exact Analysis of an Interacting {B}ose Gas. {I}.
  {T}he General Solution and the Ground State,'' Phys. Rev. {\bf 130,}
  1605--1616 (1963).

\bibitem{Lieb.63}
E.~H. Lieb, ``Exact Analysis of an Interacting {B}ose Gas. {II}. {T}he
  Excitation Spectrum,'' Phys. Rev. {\bf 130,} 1616--1624 (1963).

\bibitem{YangYang.69}
C.~N. Yang and C.~P. Yang, ``Thermodynamics of One-Dimensional Systems of
  Bosons with Repulsive Delta-Function Potential,'' J. Math. Phys. {\bf 10,}
  1115--1122 (1969).

\bibitem{Gaudin.71}
M. Gaudin, ``Boundary Energy of a {B}ose Gas in One Dimension,'' Phys. Rev. A
  {\bf 4,} 386--394 (1971).

\bibitem{Jimboetall.80}
M. Jimbo, T. Miwa, Y. M\^ori, and M. Sato, ``Density Matrix of an Impenetrable
  {B}ose Gas and the Fifth {P}ainlev\'e Transcendent,'' Physica {\bf 1D,}
  80--158 (1980).

\bibitem{Thacker.81}
H.~B. Thacker, ``Exact Integrability in Quantum Field Theory and Statistical
  Systems,'' Reviews of Modern Physics {\bf 53,} 253--285 (1981).

\bibitem{Gaudin.83}
M. Gaudin, {\em La Fonction d'Onde de {B}ethe} (Masson, Paris, 1983).

\bibitem{WhittakerWatson}
E.~T. Whittaker and G.~N. Watson, {\em A Course of Modern Analysis} (CUP,
  Cambridge, 1927).

\bibitem{PraehoferSpohn.99}
M. Praehofer and H. Spohn, ``Statistical Self-Similarity of One-Dimensional
  Growth Processes,'' cond-mat/9910273  (1999).

\bibitem{Johansson.99}
K. Johansson, ``Shape Fluctuations and Random Matrices,'' Commun. Math. Phys.
  {\bf 209,} 437 (1999).

\bibitem{KimMooreBray.91}
J.~M. Kim, M.~A. Moore, and A.~J. Bray, ``Zero-Temperature Directed Polymers in
  a Random Potential,'' Phys. Rev. A {\bf 44,} 2345--2351 (1991).

\bibitem{Halpin-Healy.91}
T. Halpin-Healy, ``Directed Polymers in Random Media: Probability
  Distributions,'' Phys. Rev. A {\bf 44,} R3415--R3418 (1991).

\bibitem{KrugMeakinHalpin-Healy.92}
J. Krug, P. Meakin, and T. Halpin-Healy, ``Amplitude Universality for Driven
  Interfaces and Directed Polymers in Random Media,'' Phys. Rev. A {\bf 45,}
  638--653 (1992).

\bibitem{Cardy.84}
J.~L. Cardy, ``Conformal Invariance and Universality in Finite Size Scaling,''
  J. Phys. A {\bf 17,} L\,385--387 (1984).

\bibitem{MMP}
J. Mathews and R.~L. Walker, {\em Mathematical Methods of Physics}
  (Addisson-Wesley, 1973).

\bibitem{HMF}
M. Abramowitz and I.~A. Stegun, {\em Handbook of Mathematical Functions} (Dover
  Publications, New-York, 1972).

\bibitem{Szego.39}
G. Szeg\"o, {\em Orthogonal Polynomials}, No.~23 in {\em Colloquium
  Publications} (AMS, New York, 1959).

\end{thebibliography}
\end{document}